\documentclass{article}

\usepackage{PRIMEarxiv}

\usepackage[utf8]{inputenc} 
\usepackage[T1]{fontenc}    
\usepackage{hyperref}       
\usepackage{url}            
\usepackage{booktabs}       
\usepackage{amsfonts}       
\usepackage{nicefrac}       
\usepackage{authblk}
\usepackage{microtype}      
\usepackage{lipsum}
\usepackage{fancyhdr}       
\usepackage{graphicx}       
\graphicspath{{media/}}     

\usepackage[longend,ruled,vlined]{algorithm2e}
\usepackage{subcaption}
\usepackage{array}
\usepackage{pgf,tikz,pgfplots}
\usepackage{float}
\usepackage{adjustbox}
\usepackage{multirow}

\newcommand{\mycomment}[1]{}

\newcommand{\JA}[1]{\textcolor{blue}{JA: #1}}

\usepackage{amsthm}       
\usepackage{amsmath}       
\usepackage{xcolor}         

\newtheorem{property}{Property}[section]
\newtheorem{lemma}{Lemma}[section]

\title{Accelerating Graph Neural Networks with a Novel Matrix Compression Format
}

\author[ ]{\textbf{João N. F. Alves}\textsuperscript{1,2,3,4}, \textbf{Samir Moustafa}\textsuperscript{1,2}, \textbf{Siegfried Benkner}\textsuperscript{1},\\ \textbf{Alexandre P. Francisco}\textsuperscript{3,4}, \textbf{Wilfried N. Gansterer}\textsuperscript{1} \& \textbf{Luís M. S. Russo}\textsuperscript{3,4}}

\affil[1]{Faculty of Computer Science, University of Vienna, Vienna, Austria}
\affil[2]{UniVie Doctoral School Computer Science, University of Vienna, Vienna, Austria}
\affil[3]{Instituto Superior Técnico de Lisboa, Universidade de Lisboa, Lisbon, Portugal}
\affil[4]{INESC-ID Lisboa, Lisbon, Portugal}

\begin{document}
\maketitle

\begin{abstract}
  The inference and training stages of Graph Neural Networks (GNNs) are often dominated by the time required to compute a long sequence of matrix multiplications between the sparse graph adjacency matrix and its embedding. To accelerate these stages, we first propose the Compressed Binary Matrix (CBM) storage format to succinctly represent the binary adjacency matrix of an unweighted graph. Then, we show how to generalize this representation to normalized adjacency matrices of unweighted graphs which arise in the context of GNNs. Finally, we develop efficient matrix multiplication kernels based on this compressed representation. The matrix multiplication kernels proposed in this work never require more scalar operations than classic sparse matrix multiplication algorithms. Experimental evaluation shows that the matrix multiplication strategies proposed outperform the current state-of-the-art implementations provided by Intel MKL, achieving speedups close to 5$\times$. Furthermore, our optimized matrix-multiplication strategies accelerated the inference time of a GNN by up to $3\times$.
\end{abstract}


\section{Introduction}

Graph Neural Networks (GNNs) are the preferred tool to learn from
graph-structured data and thus are considered key for future AI applications in
domains like social network analysis, natural language processing, biology,
physics, and many others~\cite{wu2020comprehensive}. The training and inference time of different GNN architectures is dominated by a long sequence of matrix products. This is particularly evident in GNNs that resort to Message Passing Layers (MPLs), where in each hidden layer the nodes of the graph aggregate the embedding of neighboring nodes and adjust their own embedding based on the information collected. 
In some variants of GNNs, such as the widely used Graph
Convolutional Networks (GCNs)~\cite{kipf2016semi}, the message produced in each layer is
essentially the product of the adjacency matrix of the underlying graph and its current embedding. 

For illustration purposes, consider a two-layer GCN. To propagate and combine node features across the graph, this network must compute the following operations once per inference and once per training epoch~\cite{Chai2022DistributedGN}:
\begin{equation}\label{eq:2-gcn}
    \hat{\mathbf{A}}\ \sigma(\hat{\mathbf{A}} \mathbf{X} \mathbf{W}^{0}) \mathbf{W}^{1},
\end{equation} where $\hat{\mathbf{A}}$ represents the normalized adjacency matrix of the 
graph, such that $\hat{\mathbf{A}} = \mathbf{D}^{-\frac{1}{2}} (\mathbf{A} + \mathbf{I}) \mathbf{D}^{-\frac{1}{2}}$, $\mathbf{D}$ is
the degree diagonal matrix of the graph, $\sigma$ denotes an element-wise activation function, $\mathbf{X}$ is the matrix of node features, and $\mathbf{W}^{0}$ and $\mathbf{W}^{1}$ are learnable dense matrices for the first and second layers \cite{scarselli2008graph}. It is important to note that in real scenarios the adjacency matrix of the graph is typically much larger than the remaining operand matrices of Equation~\ref{eq:2-gcn}. Therefore, matrix products involving this matrix represent most of the computational burden of training and inferring GNNs. Graphs arising in the context of GNNs are often extremely sparse. Popular GNNs frameworks, such as PyTorch~\cite{paszke2019pytorch}, leverage this sparsity to accelerate both training and inference. This is achieved by
representing the adjacency matrix of the graph, or its normalized form, in standard sparse matrix formats that only consider the nonzero elements of a sparse matrix. Sparse matrix formats, like COO or CSR, enable faster sparse-dense matrix multiplication kernels (SpMMs) which are known to require a number of scalar operations that is proportional to the number of nonzero elements in the sparse matrix.

In this paper we present a new matrix compression format called Compressed Binary Matrix (CBM) to accelerate widely-used GNN architectures (e.g., GCN \cite{kipf2016semi}, GraphSage\cite{hamilton2017inductive}, and GIN~\cite{xu2018powerful})  that spend most of their processing time on matrix products involving the adjacency matrix of the underlying graph or its normalized, when these GNNs learn on unweighted graphs. Our format exploits the fact that the adjacency matrix of an unweighted graph is binary, and therefore it can be compressed beyond what is possible with sparsity alone, while simultaneously reducing the number of scalar operation required to multiply our compressed representation of the adjacency matrix and another dense real-valued matrix.
The key advantage of the CBM format is that it only represents the differences (deltas) of a row with respect to another similar row of the same matrix, which for many use cases tends to be significantly smaller than the number of non-zeros represented by standard compression formats.

\subsection{Previous Works}\label{related-work}

The matrix-matrix and matrix-vector products have been extensively studied~\cite{Alman2018}.
A particular case is the product of a binary sparse matrix by a real-valued vector or (dense) matrix, where the efficient representation of the binary matrix can be exploited to improve both the memory footprint and the operation running time.
Although this was not expected in some preliminary studies~\cite{Karande2009}, and impossibility results exist for more complex compression schemes~\cite{Abboud2020}, it works for some representational compression schemes.
One of such schemes is the Single Tree Adjacency Forest (STAF) proposed by Nishino \emph{et al.}~\cite{Nishino2014}. The STAF representation of a binary matrix is obtained by reversing and inserting the adjacency list of each row of a binary matrix into a trie data-structure, meaning that suffixes that are common to more than one adjacency list are represented exactly once. The authors have shown that the this data-structure can be built in linear time with respect to the number of nonzero elements in the input matrix.
STAFs enable fast matrix-matrix products by traversing the trie in root-to-leaf order, while accumulating partial sums that are common to different rows of the result matrix.
The number of operations required to multiply a binary matrix represented by a STAF and a real-valued vector is proportional to the size of the trie, which is also upper-bounded by the number of nonzero elements in the input matrix.
STAFs doe not exploit, however, row-wise similarities outside of common row suffixes. To address this issue the authors proposed splitting the input matrix into sets of columns and create a STAF per set. This optimized version achieves a significant speedup and memory footprint reduction compared to CSR and the sparse matrix multiply kernel offered by the Eigen library. 

Francisco \emph{et al.}~\cite{Francisco2022} explored how succinct representations for binary matrices and graphs could be exploited to speedup binary matrix multiplication, namely the Webgraph representation by Boldi and Vigna~\cite{Boldi2004} and the Biclique Extraction (BE) based representation by Hernandez and Navarro~\cite{Hernandez2014}.
Both representations allow to reduce the memory footprint of binary matrices and accelerate the product of compressed binary matrices and real-valued vectors.
The Webgraph representation exploits the similarity among rows, as well as clustering effects present in real-world graphs and matrices, relying on gap compression, referentiation, intervalisation and $\zeta$ codes.
The BE representation exploits also rows similarities and clustering effects, extracting maximal bicliques and replacing them with differential compressed sets of nodes that share adjacencies.
The key observations are that, in both cases, we can reuse partial results from previous computations.
Taking an implementation of Pagerank using classic adjacency lists, authors achieved significant and similar speedups and memory footprint reductions, showing that the product computation can be reduced in time proportional to the compressed matrix size. 
The Webgraph and BE representations require however non-trivial preprocessing steps, Webgraph benefits from a suitable vertex reordering, obtained through graph clustering methods, and BE requires finding maximal bicliques, an NP-hard problem in general.
These steps might take longer to execute than computing the SpMM we are trying to optimize.
On the other hand the experimental evaluations considered naive implementations of standard representations, possibly leading to unfair comparisons in what concerns state of the art linear algebra libraries.

Elgohary \emph{et al.}~\cite{Elgohary2019} also addressed the problem that large-scale machine learning algorithms are often iterative, using repeated read-only data access and I/O-bound matrix-vector multiplications, introducing Compressed Linear Algebra (CLA) for lossless matrix compression.
CLA also executes linear algebra operations directly on the compressed representations.
However, it is not focused on binary matrices and, although it achieves a performance close to the uncompressed case, it only presents performance gains when data does not fit into memory.

Our work is somewhat related to the work by Bj\"{o}rklund and Lingas~\cite{BjorklundLingas2001}.
They also consider a weighted graph on the rows of a binary matrix where the weight of an edge between two rows is equal to its Hamming distance, and then they rely on a minimum spanning tree of that graph to differentially compress the rows of a binary matrix with respect to one another, thus accelerating the product computation.
They consider however only a single product of two binary matrices, and do not consider the overhead imposed by operating on a compressed representation of a binary matrix, as their results are purely theoretical.

\subsection{Our Contributions}
In this paper, we make the following contributions: \textbf{(1)}  we present the Compressed Binary Matrix (CBM) format, an efficient compression scheme for binary matrices, that can reduce the memory footprint of unweighted graphs that arise in the context of GNNs; \textbf{(2)} we introduce a new algorithm to significantly accelerate the sequence of matrix products between the (potentially normalized) adjacency matrix of the graph, represented in CBM format, and its embedding; \textbf{(3)} we prove that even in the worst-case scenario, where compression
is not possible, the number of scalar operations required to multiply a matrix represented in CBM format does not exceed the number of scalar operations required to multiply the same matrix using classic sparse storage formats; and \textbf{(4)} we have implemented the CBM format and corresponding matrix multiplication kernels such that they can be used together with state-of-the-art Deep Learning framework, such as PyTorch. Furthermore, experimental evaluation using real-world datasets demonstrates the effectiveness of our approach. Our method is nearly $5\times$ faster than state-of-the-art SpMM implementations in sequential and parallel environments, subsequently shortening the inference time of a 2-layer GCN by more than $3\times$. Our implementation will be made available at \url{https://github.com/cbm4scale}.

As previously stated, the CBM format is akin to the work of Bj\"{o}rklund and Lingas~\cite{BjorklundLingas2001}. Nevertheless, it distinguishes itself by being specifically designed to accelerate the product between a single sparse binary matrix and a set of real-valued matrices, thus the binary matrix only needs to be compressed once\footnote{Ideally, the adjacency matrix of any unweighted graph would be provided in CBM, avoiding any compression overhead.}. Furthermore, our format resorts to a Minimum Cost Arborescence (MCA), which allows us to ignore compression opportunities that do not lead to improvements in memory-footprint and matrix multiplication. The CBM format also overcomes limitations present in STAF and BE, since our format exploits compression opportunities along complete rows of the matrix (by default), is constructed in polynomial time, and does not allocate additional memory to store partial results. 

\section{Compressed Binary Matrix Format}
Let $\mathbf{A} \in \{0,1\}^{m \times n}$ be a binary matrix, where $\vec{a}_i \in \{0,1\}^{n}$ represents the $i$-th row of $\mathbf{A}$ for $i=1, \dots, m$. Also assume that $\vec{a}_i$ is represented with an adjacency list with the column indices of the nonzero elements of $\vec{a}_i$.
The Compressed Binary Matrix (CBM) format resorts to differential compression to represent the rows of a binary matrix $\mathbf{A}$. This is, if $\mathbf{A}$ is represented in CBM format, then any row $\vec{a}_x$ can be represented by another row $\vec{a}_y$ and two lists of deltas that indicate which elements must be included ($\small\Delta^{+}_{x,y}$), or removed ($\small\Delta^{-}_{x,y}$), from the adjacency list of $\vec{a}_y$ to obtain $\vec{a}_x$:
\begin{equation}\label{diff_comp}
    \vec{a}_x = (\vec{a}_y \cup \small\Delta^+_{x,y} )\setminus \small\Delta^-_{x,y}, \textrm{ where }\small\Delta^+_{x,y}=\vec{a}_x \setminus \vec{a}_y \textrm{, and } \small\Delta^-_{x,y}=\vec{a}_y \setminus \vec{a}_x.
\end{equation}

Assuming $\vec{a}_y$ is present in memory, Equation \ref{diff_comp} suggests that the memory required to represent $\vec{a}_x$ is proportional to the number of deltas between $\vec{a}_x$ and $\vec{a}_y$. If these rows are similar, then it is likely that the number of deltas is smaller than the number of nonzero elements of $\vec{a}_x$. In that case, it would be more memory efficient to represent $\vec{a}_x$ with respect to $\vec{a}_y$ than through its adjacency list.
Therefore, to further reduce the memory footprint of matrix $\textbf{A}$, the compression algorithm that builds the CBM format must find a suitable chain of compression to represent all rows of $\mathbf{A}$. This is, for each row $\vec{a}_x$, identify another similar row $\vec{a}_y$ that characterizes the former, such that: \textbf{(1)} the numbers of deltas required to represent each row $\vec{a}_x$ is minimized subjected to $\vec{a}_y$, and \textbf{(2)} the number of deltas required to represent $\Vec{a}_x$ is guaranteed to be less than, or equal to, the number of nonzero elements in $\vec{a}_x$.

\paragraph{Minimizing the number of deltas.} To address point \textbf{(1)}, the CBM format must first measure the number of deltas required to convert each row $\vec{a}_y$ into all other rows $\vec{a}_x$ of $\mathbf{A}$, i.e., measure the Hamming distance for each pair of matrix rows.
This step provides a global view of the dissimilarity between the rows of the matrix $\mathbf{A}$, and it can be modeled as a fully-connected and undirected distance graph $G$. This graph has $m$ vertices, where each vertex represents a unique row of the matrix, and the weight of each edge $(y,x)$ corresponds to the number of deltas required to represent $\vec{a}_x$ with respect to $\vec{a}_y$.
To reduce the number of deltas required to compress the rows of $\mathbf{A}$ we can find a Minimal Spanning Tree (MST) of $G$, which by definition spans $G$ with the minimum sum of edge weights possible. Naturally, any MST of the distance graph, rooted in vertex $x$, defines a chain of compression with as many deltas as the weight of this tree plus the number of nonzero elements of $\vec{a}_x$. Thus, any MST rooted in the vertex corresponding to the row with the fewest nonzero elements, defines a chain of compression that satisfies point \textbf{(1)}.

\paragraph{Worst-case guarantees.} Note that the chain of compression obtained by finding an MST of $G$ does not satisfy point \textbf{(2)}, because the weight of the lightest incoming edge of any vertex $x$ can be greater than the number of nonzero elements in $\vec{a}_x$. In such cases, representing $\vec{a}_x$ with an adjacency list is clearly more memory efficient. To avoid this issue, we extended the distance graph $G$ with a virtual vertex $\mathbf{0}$ which is connected to all other vertices of the graph. This virtual vertex represents a null row-vector $\vec{a}_0 \in\{0\}^n$, which ensures that the weight of each edge $(\mathbf{0},x)$ is equal to the number of nonzero elements in $\vec{a}_x$. The inclusion of virtual vertex $\mathbf{0}$ in the distance graph $G$ ensures that the issue described above cannot occur, since the lightest incoming edge of any vertex $x$ is now at most as heavy as the number of nonzero elements in $\vec{a}_x$. Therefore, any chain of compression characterized by an MST of $G$, rooted on vertex $\mathbf{0}$, is guaranteed to satisfy points \textbf{(1)} and \textbf{(2)}. If we use this chain of compression to represent a matrix in CBM format, then following property will be observed:

\begin{property}\label{prop:1}
    The number of deltas required to represent any matrix $\mathbf{A}\in\{0,1\}^{m \times n}$ in Compressed Binary Matrix (CBM) format is never greater than the number of nonzero elements in matrix $\mathbf{A}$.
\end{property}

To complete the construction, we simply traverse the compression chain above in topological order, and for every edge $(y,x)$ visited, we compute the lists of positive and negative deltas required to convert row $\vec{a}_y$ into $\vec{a}_x$.

\subsection{Time and Space Analysis}\label{timespace}
\begin{lemma}\label{lemma1}
Any matrix $\mathbf{A}\in\{0,1\}^{m \times n}$ can be represented in CBM format in $O((m+1)\ \mathbf{nnz}(\mathbf{A}) + m^2 \log m)$ time. 
\begin{proof}
    The construction of the extended distance graph $G$ for $\mathbf{A}$ requires the computation of $m(m+1)/2$ Hamming distances between all possible row pairs $(\vec{a}_x, \vec{a}_y)$. The Hamming distance of each pair of rows can be reduced to the intersection of their adjacency lists, computed in $O(\mathbf{nnz}(\vec{a}_x) + \mathbf{nnz}(\vec{a}_y))$ time. Hence, the time to compute all $m(m+1)/2$ Hamming distances is upper-bounded by
    \begin{equation}
        \sum_{x=0}^m \sum_{y=0}^m \big(\mathbf{nnz}(\vec{a}_x) + \mathbf{nnz}(\vec{a}_y)\big) = (m+1)\ \mathbf{nnz}(\mathbf{A}).
    \end{equation}
Additionally, well-known MST algorithms, such as Prim or Kruskal, are known to find an MST in $O(E \log V)$ time, where $E$ and $V$ denote the number of edges and vertices in the graph. Since the extended distance graph $G$ contains $m(m+1)/2$ edges and $m+1$ vertices, finding an MST of $G$ requires $O((m+1)^2 \log (m+1))$ time, and therefore, representing matrix $\mathbf{A}$ takes time
\begin{equation}\label{time_cbm}
    O((m+1)\ \mathbf{nnz}(\mathbf{A}) + (m+1)^2 \log (m+1))).
\end{equation}
Finally, a single list of deltas can also be obtained from the intersection of the adjacency lists of $\vec{a}_x$ and $\Vec{a}_y$. Hence, the computation of $2m$ lists of deltas is already accounted for in Equation \ref{time_cbm}. 
\end{proof}
\end{lemma}

\mycomment{
The CBM representation of a binary matrix $A$ is fully characterized by the edges of an MST of the extended distance graph $G$, and the lists of positive ($\small\Delta^+$) and negative ($\small\Delta^-$) deltas associated with each vertex of $G$. Therefore, the memory required to represent a matrix in CBM format depends on the number of rows in $A$ and the sum of the size of all lists of deltas.
}

\begin{lemma}\label{cbm:space}
The space required to represent a binary matrix $\mathbf{A}\in\{0,1\}^{m \times n}$ in Compressed Binary Matrix (CBM) format is proportional to $O(m + \sum_{i=0}^{m} (|\small\Delta^{+}_{x, r_x}| + |\small\Delta^{-}_{x, r_x}|))$, where $r_x$ represents the index of the row selected to compress row $\vec{a}_x$.
\vspace{-1ex}
    \begin{proof}
        Assuming matrix $\mathbf{A}$ is represented in CBM format. Then the chain of compression required to represent this matrix consists on: \textbf{(1)} a list of edges that represents any MST rooted in vertex $\mathbf{0}$ of graph $G$, and \textbf{(2)} a list of positive and negative deltas for each edge of this MST. Since the extended version of $G$ is a fully-connected graph with $m+1$ vertices, it is known that any MST of $G$ contains $m$ edges. Therefore, the list of edges contains $m$ elements, and there are $m$ lists of deltas, whose size totals $\sum_{x=0}^{m} (|\small\Delta^{+}_{x,r_x}| + |\small\Delta^{-}_{x,r_x}|)$.
    \end{proof}
\end{lemma}

\subsection{Fast Matrix Multiplication with Compressed Binary Matrix Format}\label{spmm_cbm}

Let $\Vec{w}\in \mathbb{R}^n$ be a dense and real-valued vector, and $\vec{a}_x$ and $\vec{a}_y$ two distinct rows of a binary matrix $\mathbf{A}$ as previously defined. 
It follows from Equation \ref{diff_comp} that we can resort to the inner-product $\vec{a}_y \cdot \vec{w}$ to compute $\vec{a}_x \cdot \vec{w}$ as
\begin{equation}\label{diff_mul}
   \vec{a}_x \cdot \vec{w}  = ((\vec{a}_y \cup \small\Delta^+_{x,y} )\setminus \small\Delta^-_{x,y})\cdot\vec{w} = (\vec{a}_y \cdot \vec{w}) + (\small\Delta^+_{x,y}\cdot\vec{w}) - (\small\Delta^-_{x,y}\cdot\vec{w}). 
\end{equation}
This implies that the dot-product $\vec{a}_x \cdot \vec{w}$ can be calculated in $1+|\small\Delta^+_{x,y}|+|\small\Delta^-_{x,y}|$ scalar operations, if the value of $\Vec{a_y} \cdot \vec{w}$ can be reused. Naturally, we can resort to this strategy to design fast matrix-vector multiplication kernels $\vec{u}\gets\mathbf{A}\vec{v}$, where $\mathbf{A}\in\{0,1\}^{m \times n}$, $\Vec{u}\in\mathbb{R}^m$, and $\Vec{v}\in\mathbb{R}^n$, by computing all dot-products between the rows of $\mathbf{A}$ and $\Vec{v}$ in an order where: \textbf{(1)} the value of the dot-product $\vec{a}_y \cdot \Vec{v}$ is known before calculating $\vec{a}_x \cdot \Vec{v}$, and \textbf{(2)} the value of all dot-products $\vec{a}_x \cdot \vec{v}$ is calculated with respect to the value $\vec{a}_y \cdot \vec{v}$ that results in the minimum overall number of scalar operations. By definition, the chain of compression of matrix $\mathbf{A}$ already represents such an ordering. Therefore, we can accelerate matrix-vector multiplication by traversing the chain of compression of matrix $\mathbf{A}$ in topological order, and for each edge visited compute $u_x\gets \Vec{a}_x \cdot \Vec{v}$ as
\begin{equation}\label{eq:5}
  u_x \gets u_y + (\small\Delta^{+}_{x,y} \cdot \vec{v}) - (\Delta^{-}_{x,y} \cdot \vec{v}),
\end{equation}
where $u_y$ is known to already contain the value $\Vec{a}_y \cdot \Vec{v}$. Note that classic sparse-dense matrix-vector multiplication kernels compute $\Vec{u}\gets \mathbf{A} \cdot \Vec{v}$ in $\mathbf{nnz}(\mathbf{A})$ scalar operations, where $\mathbf{nnz}(\mathbf{A})$ represents the numbers of nonzero elements in matrix $\mathbf{A}$. Assuming $\mathbf{A}$ is represented in CBM format, then the number of deltas required to represent each row $\vec{a}_x$ of $\mathbf{A}$ is known to be smaller than, or equal to, the number of nonzero elements in $\vec{a}_x$. If the number of deltas required to represent $\vec{a}_x$ is strictly smaller than $\mathbf{nnz}(\vec{a}_x)$, then it is clear that the dot-product $\vec{a}_x\cdot\vec{v}$ requires at most $\mathbf{nnz}(\vec{a}_x)$ scalar operations. On the other hand, if the number of deltas required to represent $\vec{a}_x$ is the same as the number of nonzero elements in $\vec{a}_x$, Equation \ref{diff_mul} suggests that the dot-product $\vec{a}_x \cdot \vec{v}$ would be computed in $1+\mathbf{nnz}(\vec{a}_x)$ scalar operations. This scenario can be avoided by engineering the MST algorithm to select an out-going edge of the virtual node $\textbf{0}$ in case of draw. Since $\vec{a}_x$ was not compressed in this scenario, the number of scalar operations required to compute this dot-product is exactly $\mathbf{nnz}(\vec{a}_x)$.\mycomment{  \JA{comment: This next part is confusing. Maybe we could just say that if the nnz of $a_x$ is equal to the deltas then this edge is not considered during the construction of the format}However, if the number of deltas required to represent row $\vec{a}_x$ is the same as the number of nonzero elements in $\vec{a}_x$, then this row can be compressed with respect to the null-row $\vec{a}_0$. This implies that the dot-product $\vec{a}_x \cdot \vec{v}$ can be computed in exactly $\mathbf{nnz}(\vec{a}_x)$, since $\vec{a}_0\cdot\Vec{v}=0$, meaning that the cost of this addition can be engineered away.} As the number of scalar operations required to compute $\Vec{a}_x \cdot \Vec{v}$ never surpasses the number of nonzero elements in $\vec{a}_x$ for $x=1,\dots,m$, the following property becomes evident:
\begin{property}\label{prop:2}
    The number of scalar operations required to compute matrix-vector multiplication based on the Compressed Binary Matrix (CBM) format is never greater than those required to compute matrix-vector multiplication based on classic sparse formats.
\end{property}
Additionally, Equation \ref{diff_mul} shows that our matrix-vector multiplication strategy does not require the allocation of additional buffers, since the value of the dot-product required to compute $\Vec{a}_x \cdot \Vec{v}$ is guaranteed to already be stored in vector $\vec{u}$. Therefore, the following property is observed:
\begin{property}\label{prop:3}
    The amount of memory required to compute matrix-vector multiplication based on the Compressed Binary Matrix (CBM) format is proportional to the size of its operands and remains constant during execution time. 
\end{property}

Intuitively, we can resort to the matrix-vector multiplication strategy described above to design fast matrix-matrix multiplication kernels as described in Algorithm \ref{alg:mm}. This algorithm assumes that the left-hand side operand matrix $\mathbf{A}\in\{0,1\}^{m\times n}$ is represented in CBM format, while matrices $\mathbf{B}\in\mathbb{R}^{n \times p}$ and $\mathbf{C}\in\mathbb{R}^{m\times p}$ are dense and correspond to the right-hand side operand matrix and to the product matrix, respectively. As it can be observed, Algorithm \ref{alg:mm} computes the matrix-vector product between matrix $\mathbf{A}$ and each column of matrix $\mathbf{B}$. Therefore, we can conclude that Properties \ref{prop:2} and \ref{prop:3} hold for the matrix-matrix multiplication case.
\begin{algorithm}[b]
    \caption{Matrix-Matrix product with CBM}\label{alg:mm}
    \KwIn{$\mathbf{A} \in \{0,1\}^{m\times n}$, $\mathbf{B} \in \mathbb{R}^{n\times p}$, $\mathbf{C} \in \mathbb{R}^{m\times p}$} 
    \PrintSemicolon
    \BlankLine
    
    \nl\ForEach{\textup{edge} $(y, x)\in \text{ chain of compression of } \mathbf{A}$\textup{ (in topological order)}}
    {
        \nl\ForEach{\textup{column vector} $\vec{b}_k \in \mathbf{B}$ \textup{, where} $k\in [1,p[$ }
        {
                \nl$\mathbf{C}_{x, k} \gets \mathbf{C}_{y, k} + (\small\Delta^{+}_{x,y} \cdot \vec{b}_k) - (\small\Delta^{-}_{x,y} \cdot \vec{b}_k)$\;
        }
    }
    \Return $\mathbf{C}$
\end{algorithm}
\paragraph{Leveraging High Performance SpMM kernels. } The representation of a binary matrix $\mathbf{A}$ in CBM format was until now conceptualized as a chain of compression, where each node of this chain is associated with two lists of deltas. As is, a compressed matrix cannot be represented in a sparse format capable of leveraging efficient SpMM kernels. To address this issue, we represent the lists of deltas that characterize our format as a matrix $\mathbf{A'}\in\mathbb{R}^{m \times n}$, which can be represented in any convenient matrix format, and leverage efficient SpMM kernels provided by Intel MKL to compute all dot-products of Algorithm \ref{alg:mm} in a single matrix product $\mathbf{A'}\mathbf{B}$. Once $\mathbf{A'}\mathbf{B}$ is stored in matrix $\mathbf{C}$, we can finalize the matrix multiplication with CBM, by traversing the chain of compression in topological order, and updating row $\vec{c}_x$ of $\mathbf{C}$ as $\vec{c}_x := \vec{c}_x + \vec{c}_y$ for each edge $(y,x)$ that was visited. Naturally, these updates can also leverage efficient AXPY kernels, which are also provided by Intel MKL.

\mycomment{\paragraph{Leveraging High Performance SpMM kernels. } The representation of a binary matrix $\mathbf{A}$ in CBM format was until now conceptualized as a chain of compression, where each node of this chain is associated with two lists of deltas. As is, a compressed matrix cannot be represented in a sparse format capable of leveraging efficient SpMM kernels. To address this issue, we represent the lists of deltas that characterize our format as a matrix $\mathbf{A'}\in\mathbb{R}^{m \times n}$, which can be represented in any convenient matrix format, and leverage efficient SpMM kernels provided by Intel MKL to compute all dot-products of Algorithm \ref{alg:mm} in a single matrix product $\mathbf{A'}\mathbf{B}$. Once $\mathbf{A'}\mathbf{B}$ is stored in matrix $\mathbf{C}$, we can finalize the matrix multiplication with CBM, by traversing the chain of compression in topological order, and accordingly updating the value of row $\vec{c}_x$ of $\mathbf{C}$ with the current value of $\vec{c}_y$. Naturally, these updates can also leverage efficient AXPY kernels, which are also provided by Intel MKL.}

\paragraph{Multi-threading parallelism.} As suggested in the previous paragraph, matrix multiplication with the CBM format can be divided into two stages. The first one computes the product $\mathbf{C}:=\mathbf{A'B}$ which is embarrassingly parallel, and Intel MKL already provides efficient multi-threaded and vectorized implementations of this operation. The second stage involves updating the rows of matrix $\mathbf{C}$ with respect to the chain of compression. This stage presents data-dependencies, since the final value of row $\vec{c}_x$ of $\mathbf{C}$ can only be calculated once $\vec{c}_y$ is known. There are however no dependencies between different branches of the chain of compression. Therefore, we can parallelize this stage 
by concurrently updating the rows of matrix $\mathbf{C}$ that are found in different branches of the virtual node $\textbf{0}$.

\paragraph{Extending SpMM with CBM to normalized adjacency matrices.} Let $\hat{\mathbf{A}}\in\mathbb{R}^{n \times n}$ represent the normalized adjacency matrix of an unweighted graph, where $\hat{\mathbf{A}} = \mathbf{D}^{-\frac{1}{2}} (\mathbf{A} + \mathbf{I}) \mathbf{D}^{-\frac{1}{2}}$. As is, we cannot resort to the CBM format to represent $\hat{\mathbf{A}}$ since this matrix is not binary. However, fast matrix multiplication is still possible. Note that $\mathbf{D}^{-\frac{1}{2}}$ is a diagonal matrix, and therefore $(\mathbf{A} + \mathbf{I}) \mathbf{D}^{-\frac{1}{2}}$ corresponds to a column-scaled matrix, i.e., the elements in a column of the matrix are either 0 or have a constant value that is unique to this column. We can represent this matrix in CBM format, by simply multiplying the corresponding matrix of deltas $(\mathbf{A'} + \mathbf{I})$ by $\mathbf{D}^{-\frac{1}{2}}$. At this point, we can efficiently compute $\mathbf{C}:=((\mathbf{A} + \mathbf{I})\mathbf{D}^{-\frac{1}{2}})\mathbf{B}$ as described previously. 
Finally, note that $\mathbf{D}^{-\frac{1}{2}}(\mathbf{A} + \mathbf{I})\mathbf{D}^{-\frac{1}{2}}\mathbf{B})$, simply scales the rows of matrix $(\mathbf{A} + \mathbf{I})\mathbf{D}^{-\frac{1}{2}}\mathbf{B}$. The cost of scaling the rows of this matrix can be hidden by fusing it with the update step of the product matrix $\mathbf{C}$. 

\paragraph{Speeding up SpMM with Edge Pruning.} Note that not all compression opportunities contribute to faster matrix multiplication kernels. The overheads associated with differential compression, such as traversing the chain of compression and updating the result matrix, might overcome potential performance gains if the number of scalar operations saved are not above a certain threshold. To address this issue, we can prune all edges of the distance graph of $\mathbf{A}$ where the number of scalar operations saved does not meet a user-defined threshold $\alpha\in\mathbb{N}$. For each edge $(y,x)$ in the extended distance graph of $\mathbf{A}$, prune this edge if its weight is greater than the number of nonzero elements in $\vec{a}_x$ minus $\alpha$. Naturally, if we prune the edges of the extended distance graph of $\mathbf{A}$ in this manner, it is possible for a single edge direction to be pruned, while the opposite direction remains in the graph. Therefore, the extended distance graph of $\mathbf{A}$ is now directed, and a suitable chain of compression corresponds to a Minimum Cost Arborescence (MCA) rooted in the virtual node $\textbf{0}$. Note that our compression algorithm remains correct, since the extended distance graph of $\mathbf{A}$ contains an out-going edge from the virtual node $\textbf{0}$ to all other nodes; Furthermore, the time required to build the CBM format, as shown in Lemma \ref{lemma1}, remains unchanged since finding an MCA or an MST present the same time complexity ~\cite{bother2023efficiently}.

\section{Experimental Evaluation}
\paragraph{Experimental Setting.} The experiments found in this section were run on an Intel Xeon Gold 6130 (Skylake) CPU with 16 physical cores and 2.1 GHz fixed clock frequency. This machine runs on CentOS Linux 7 (version 3.10.0) operating system. The SpMM kernels tested in this section were implemented in C++, and rely on Intel MKL 24.0 sparse CSR format and corresponding matrix multiplication kernels. The C++ code developed in this work is then called from Python 3.11, via PyTorch C++ Extensions to reliably emulate common use-cases. 
Parallel experiments (with 16 cores) were implemented with OpenMP 4.5, and the threads were pinned to physical cores with \texttt{GOMP\_CPU\_AFFINITY="0-15"} environment variable.

    
    


\subsection{Evaluation Metrics}
The CBM format was evaluated with respect to quality of compression achieved ({\em compression ratio}), and the time required ({\em runtime reduction}) to compute SpMM and to infer a 2-layer GCN, when the adjacency
matrix of the graphs are represented in our format. The {\em compression
ratio} is defined as the ratio between the memory required to represent a matrix in CSR format and 
CBM format. In our implementation, the CBM format is composed by the corresponding matrix of deltas $\mathbf{A'}$ and a tree representing the chain of compression, both stored in CSR format. In the context of sparse-dense matrix multiplication (SpMM), the {\em runtime reduction} is measured by comparing the average time, out of
50 runs, it takes to perform matrix multiplication with a randomly generated
dense matrix with 500 columns using the CSR format, to the time taken to compute
the same matrix multiplication with the CBM format. The formula used to capture
this metric is $\frac{T_{\text{CSR}} - T_{\text{CBM}}}{T_{\text{CSR}}} \times
100\%$, where $T_{\text{CSR}}$ is the time required to carry out sparse-dense matrix multiplication with CSR by the state-of-the-art SpMM implementation offered by Intel MKL, and $T_{\text{CBM}}$ is the time taken to compute the same product with CBM. We used same formula and number of runs in the context of GCN inference, however, $T_{\text{CSR}}$ and $T_{\text{CBM}}$ correspond to the time required by the inference stage of this network by resorting to SpMM kernels based on CSR and CBM, respectively. It is important to note that we did not consider the SpMM kernels that are native to PyTorch, because these kernels were substantially slower than the ones implemented in Intel MKL.  

\subsection{Datasets}

To demonstrate the advantages of the CBM format in the context of SpMM and GCN inference, we selected six real-world graphs of varying size and average in-degree as depicted in Table \ref{tab:datasets}. The selected graphs depict relationships between authors and/or academic papers, where nodes tend to share many common neighbors. This property suggests that the adjacency matrices of these graphs are good candidates to be represented in CBM format. 

In co-paper graphs, each node represents a paper. An undirected edge is placed between two nodes if the corresponding papers share at least one common author. They depict the interconnection and collaborative patterns between various academic publications.
Co-author graphs represent scientific collaborations between authors of academic papers. Here, nodes correspond to authors, and an undirected edge is placed between two nodes if authors of the nodes have co-authored a paper together. If a paper is authored collaboratively by a group of authors, it results in a fully connected subgraph, or clique, encompassing those grouped nodes.
Citation graphs are directed graphs where each node represents an academic paper. Directed edges in these graphs illustrate citations, with an edge pointing from the citing paper to the cited paper. These graphs highlight the directional flow of information and the influence of one paper upon another within the academic community.

\begin{table}[h]
\caption{Summary of key datasets used to evaluate SpMM and GNN efficiency and scalability, categorized by graph type and sorted by average in-degree. This table presents the total number of nodes, edges, and average in-degree.}
\label{tab:datasets}
\centering
\begin{tabular}{l l r r r}
\toprule
\textbf{Graph} & \textbf{Type} & \textbf{Nodes} & \textbf{Edges} & \textbf{Avg. In-degree} \\
\midrule
coPapersCiteseer~\cite{Rossi2015TheND} & Co-paper          & 434,102  & 32,073,440 & 74.8 \\
coPapersDBLP~\cite{Rossi2015TheND}     & Co-paper          & 540,486  & 30,491,458 & 57.4 \\
ca-AstroPh~\cite{snapnets}          & Co-author         & 18,772   & 396,160    & 22.1 \\
ca-HepPh~\cite{snapnets}        & Co-author         & 12,008   & 237,010    & 20.7 \\
PubMed~\cite{Yang2016RevisitingSL}              & Citation network  & 19,717   & 88,648     & 5.4  \\
Cora~\cite{Yang2016RevisitingSL}           & Citation network  & 2,708    & 10,556     & 4.8  \\
\bottomrule
\end{tabular}
\end{table}

\subsection{Sparse-Dense Matrix Multiplication (SpMM) Evaluation} \label{spmm}
Finding the best $\alpha$ is key to improve the performance of matrix multiplication with CBM. Adjusting this parameter not only reduces overhead associated with
traversing the compression chain, but also exposes more parallelism opportunities as it increases the out-degree of the virtual node. Given the importance of $\alpha$ we first consider the case where $\alpha = 0$ and our edge pruning technique was not applied, and then we show how fine-tuning $\alpha$ improves the quality of matrix multiplication with CBM.   
%
\begin{figure}[h]
    \centering
    \includegraphics[width=\linewidth]{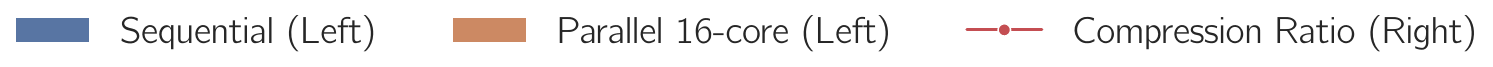}

    \begin{subfigure}{0.333\textwidth}
        \centering
        \includegraphics[width=\linewidth]{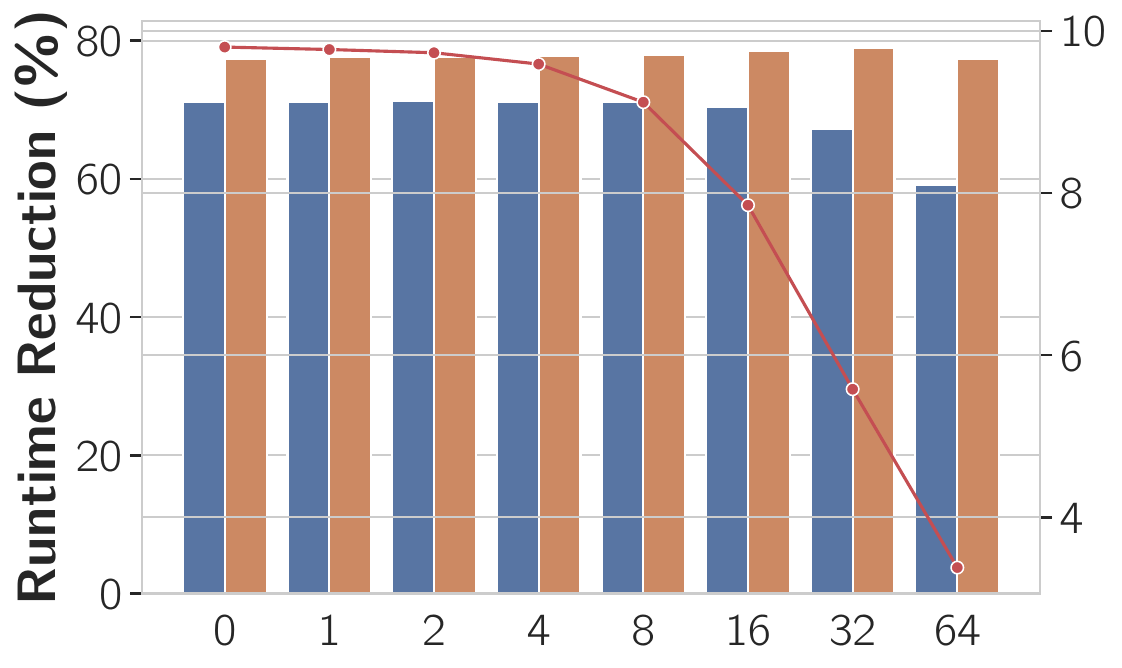}
        \caption{coPapersCiteseer Dataset}
        \label{citeseer_mm}
    \end{subfigure}\hfill
    \begin{subfigure}{0.333\textwidth}
        \centering
        \includegraphics[width=\linewidth]{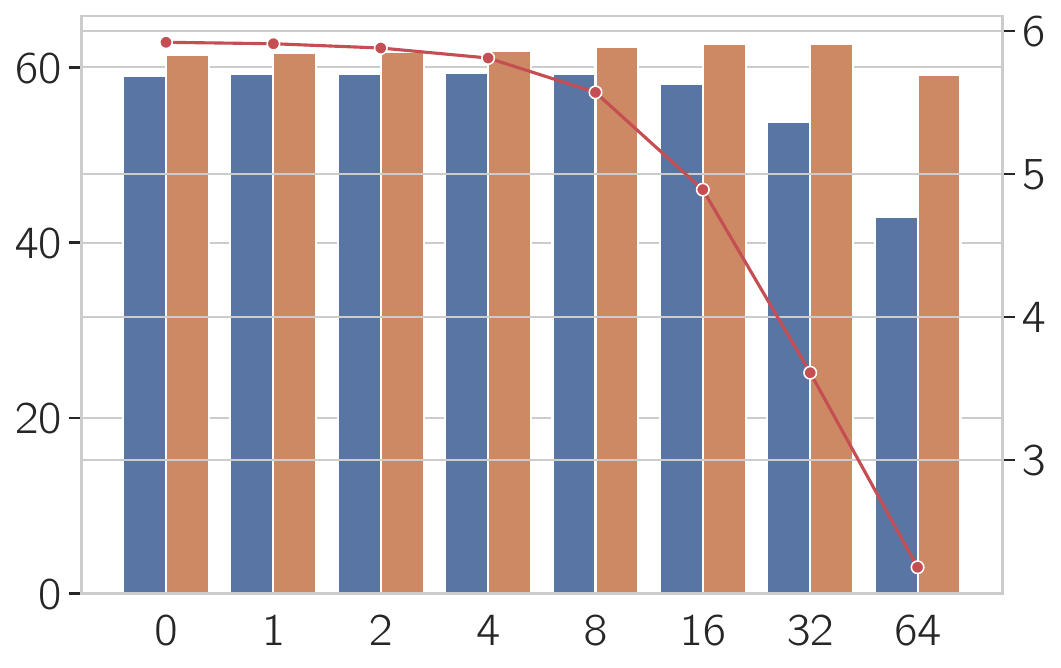}
        \caption{coPapersDBLP Dataset}
        \label{dblp_mm}
    \end{subfigure}\hfill
    \begin{subfigure}{0.333\textwidth}
        \centering
        \includegraphics[width=\linewidth]{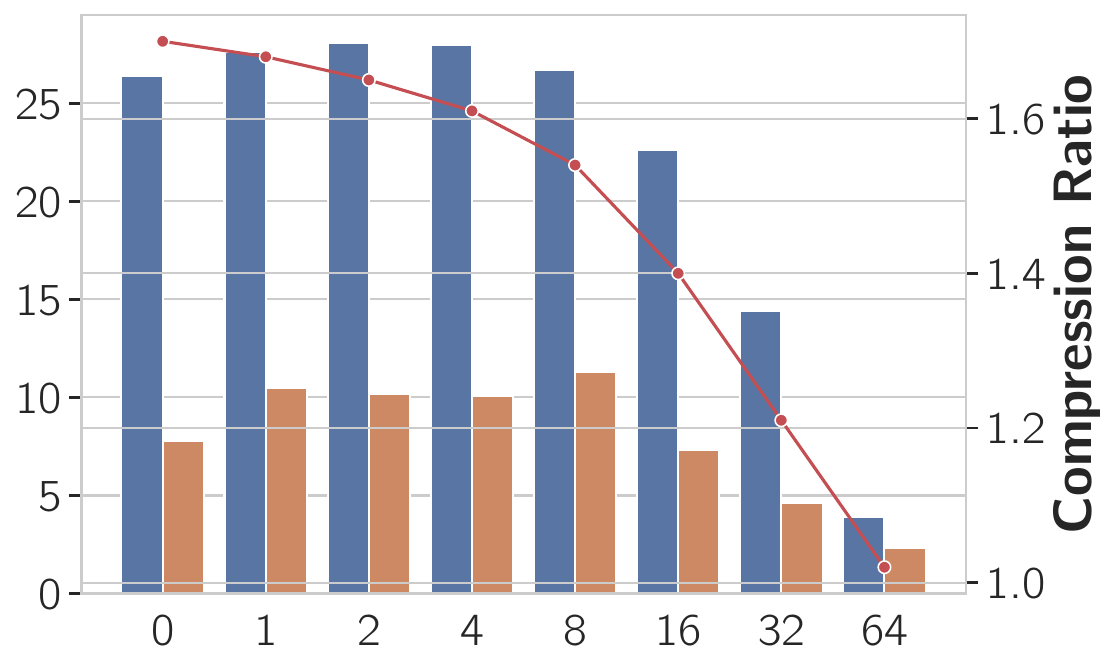}
        \caption{ca-AstroPh Dataset}
        \label{astro_mm}
    \end{subfigure}

    \begin{subfigure}{0.333\textwidth}
        \centering
        \includegraphics[width=\linewidth]{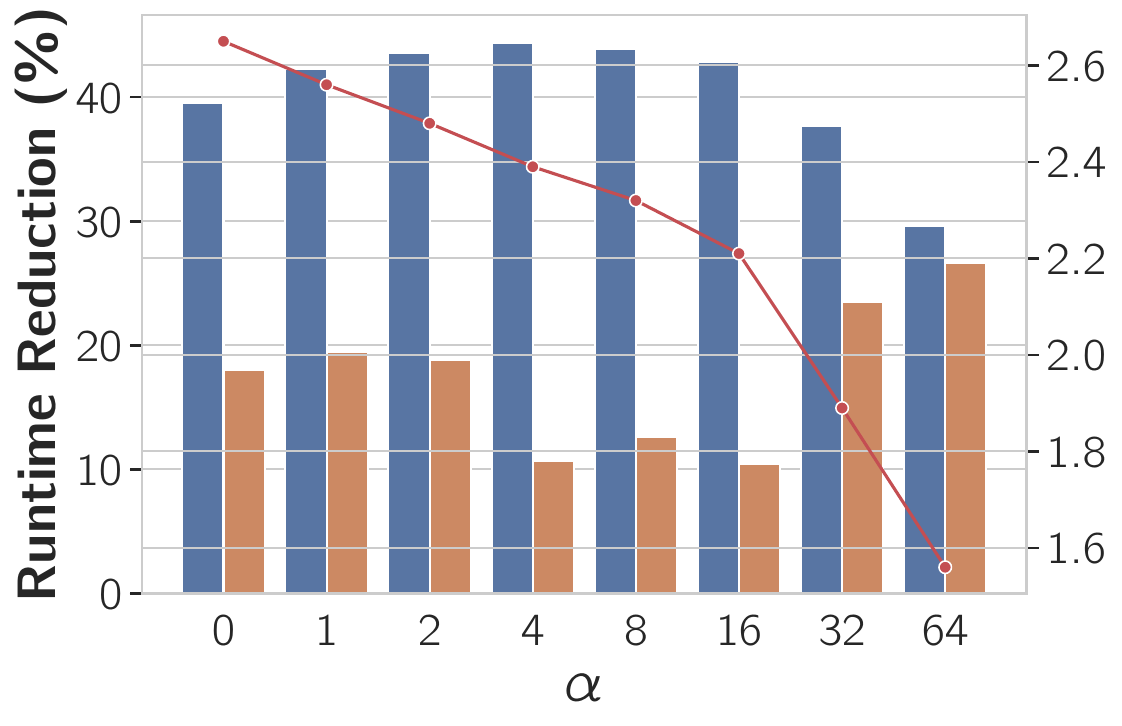}
        \caption{ca-HepPh Dataset}
        \label{hep_mm}
    \end{subfigure}\hfill
    \begin{subfigure}{0.333\textwidth}
        \centering
        \includegraphics[width=\linewidth]{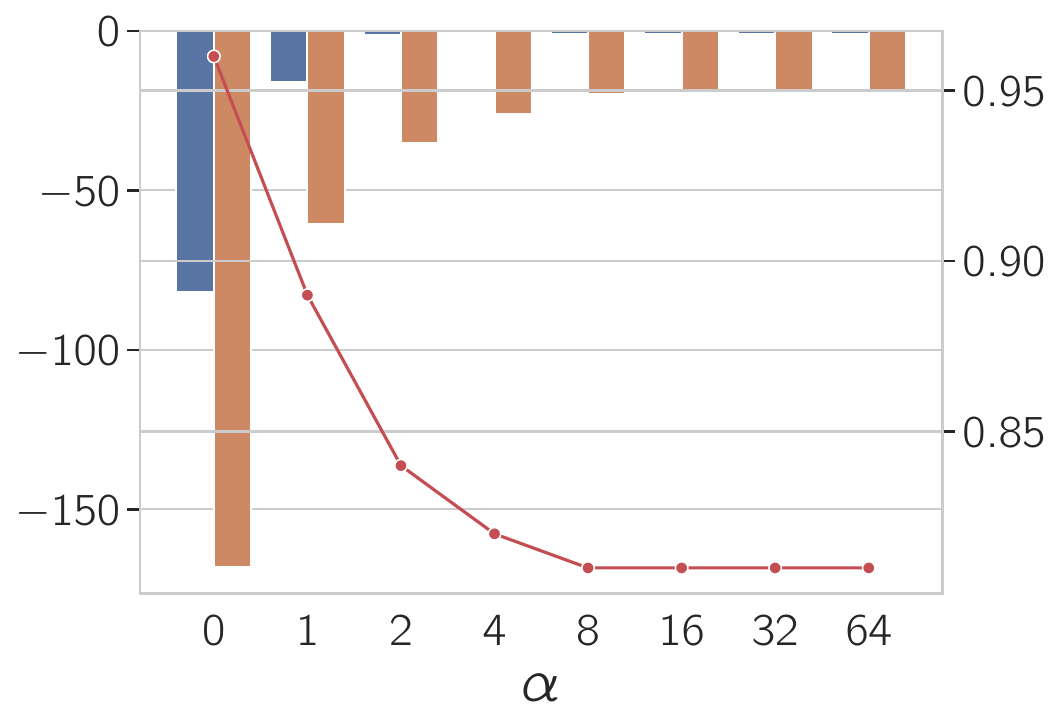}
        \caption{Cora Dataset}
        \label{cora_mm}
    \end{subfigure}\hfill
    \begin{subfigure}{0.333\textwidth}
        \centering
        \includegraphics[width=\linewidth]{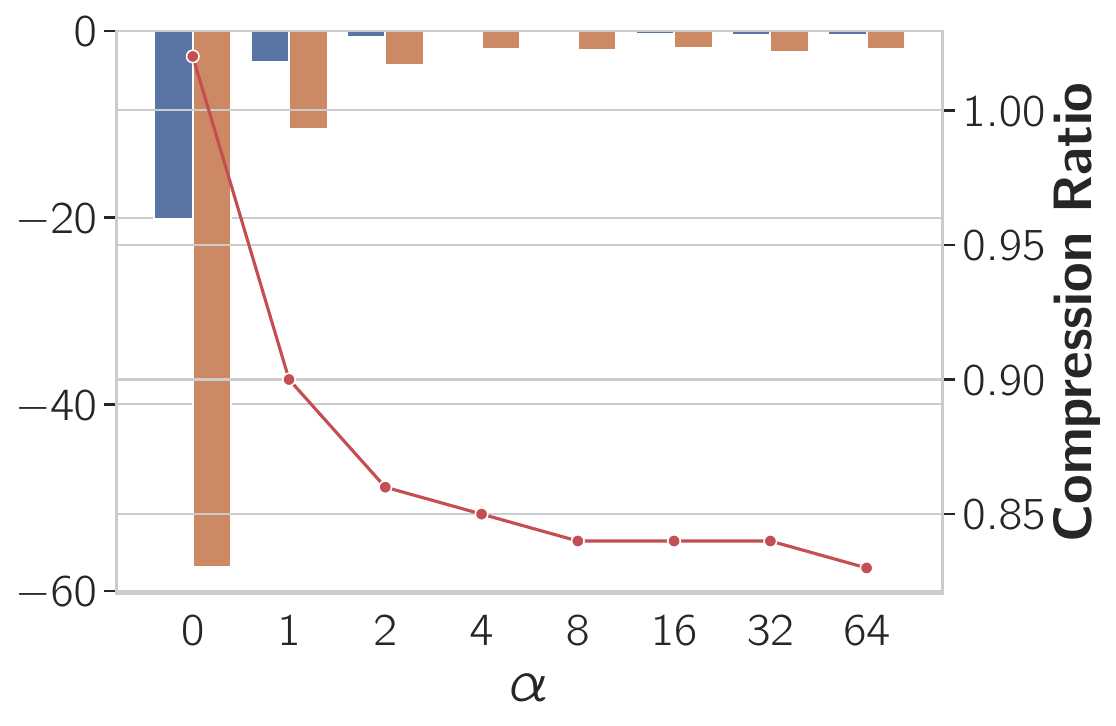}
        \caption{PubMed Dataset}
        \label{pub_mm}
    \end{subfigure}

    \caption{Performance impact of different $\alpha$ values for the CBM format. These plots compare runtime reduction in sequential and parallel environments, along side with the compression rate across various $\alpha$ values. The x-axis lists the $\alpha$ values, the left y-axis shows runtime reduction relative to the original sparse-dense matrix product (SpMM), while the right y-axis shows the compression ratio against the dataset size in CSR.}
    \label{fig:spmm_mkl}
\end{figure}


\mycomment{impacts the The behaviors of $\alpha$ for each datasets when setting the $\alpha$ parameter to different values, impacting both compression effectiveness and computational performance. \JA{This paragraph is very confusing. Please be direct, no point on trying to be fancy}}

\paragraph{$\alpha$ equal to zero.} Representing Cora (Fig. \ref{cora_mm}) and PubMed (Fig. \ref{pub_mm}) datasets in CBM resulted in minimal, and even negative, compression
gains with respect to CSR. This is likely caused by the small average in-degree of these graphs, which suggests that the compression opportunities found in these graphs do not offset the memory overhead required to represent the corresponding chains of compression. As expected, the poor compression rate of these datasets led to no speedup in the context of matrix multiplication. 
On the other hand, compressing ca-AstroPh (Fig. \ref{astro_mm}) and ca-HepPH (Fig. \ref{hep_mm}) with our format respectively increased the compression ratios of both datasets to 2.6$\times$ and 1.7$\times$, subsequently accelerating matrix product for both datasets. Our matrix multiplication strategy achieved a runtime reduction of 26\% and 8\% for ca-AstroPh in sequential and parallel environments, respectively. For ca-HepPh the same multiplication kernel presented a runtime reduction of 40\% and 18\% also in sequential and parallel environments. 
The datasets that benefited the most from our format were the coPapersCiteseer (Fig. \ref{citeseer_mm}) and coPapersDBLP (Fig. \ref{dblp_mm}), achieving impressive compression ratios of 9.8$\times$ and 5.9$\times$, leading to substantial improvements in computational performance. Our matrix multiplication kernel with coPapersCiteseer achieved a runtime reduction of 71\% and 77\% in sequential and parallel environments, while the same kernel with coPapersDBLP presented a runtime reduction of 59\% and 61\% in the same experimental settings. These results highlight the potential of the CBM format to efficiently compress and multiply unweighted graphs that present natural communities and an high average in-degree.
%

\paragraph{$\alpha$ greater than zero.} Setting $\alpha$ to a value greater than 8 reduced the overhead associated with traversing the chain of compression of Cora (Fig. \ref{cora_mm}) and PubMed (Fig. \ref{pub_mm}), making the performance decay observed for $\alpha = 0$ negligible in sequential and parallel settings, even when no compression gains are observed. 
Increasing the value of $\alpha$ is also beneficial for sequential matrix multiplication with ca-AstroPh (Fig. \ref{astro_mm}) and ca-HepPh (Fig. \ref{hep_mm}), improving the respective runtime reduction from 26\% up to 28\% ($\alpha=2$) and from 40\% up to 44\% ($\alpha=4$). Nevertheless, it is important to note that our compression algorithm will start to ignore good compression opportunities once $\alpha$ is large enough, worsening the performance of our matrix multiplication strategy. This effect is evident in both datasets for $\alpha$ greater than 16, where the compression ratio decreases alongside with the runtime reduction. 
Experiments with both co-author datasets show that adjusting $\alpha$ is even more important in the parallel case, increasing the runtime reduction of our multiplication strategy with ca-AstroPh from 8\% up to 11\% ($\alpha=8$), and with ca-HepPh from 18\% up to 26\% ($\alpha=64$). Furthermore, these experiments confirm that the degree of parallelism of our multiplication strategy increases concurrently with $\alpha$. This effect is easily observed for ca-HepPh (Fig.\ref{hep_mm}), where the performance of our matrix multiplication kernel sharply declines for $\alpha$ greater than 2, followed by a steep increase in performance when $\alpha$ equals to 32 or 64 (even though our compression algorithm is already ignoring good compression opportunities at this point). 
Finding the best $\alpha$ is not as relevant for coPapersCiteseer (Fig.\ref{citeseer_mm})  and coPapersDBLP (Fig.\ref{dblp_mm}), as most compression opportunities save more than 8 scalar operations in the context of our matrix multiplication strategy. This observation is verified, since the compression ratio of these datasets shows little to no decrease for $\alpha$ smaller than 8. This effect is most likely due to the high average in-degree of both datasets. Still, adjusting $\alpha$ is required to obtain the best runtime reduction for both coPapers datasets. In our experiments our matrix multiplication kernel with coPapersCiteseer achieved a peak runtime reduction of 71\% ($\alpha=2$) and 79\% ($\alpha=32$) in sequential and parallel environments, while the same kernel with coPapersDBLP peaks at 59\% ($\alpha=$4) and 63\% ($\alpha=16$) also in sequential and parallel environments.

\begin{figure}[b]
    \centering
    \includegraphics[width=\linewidth]{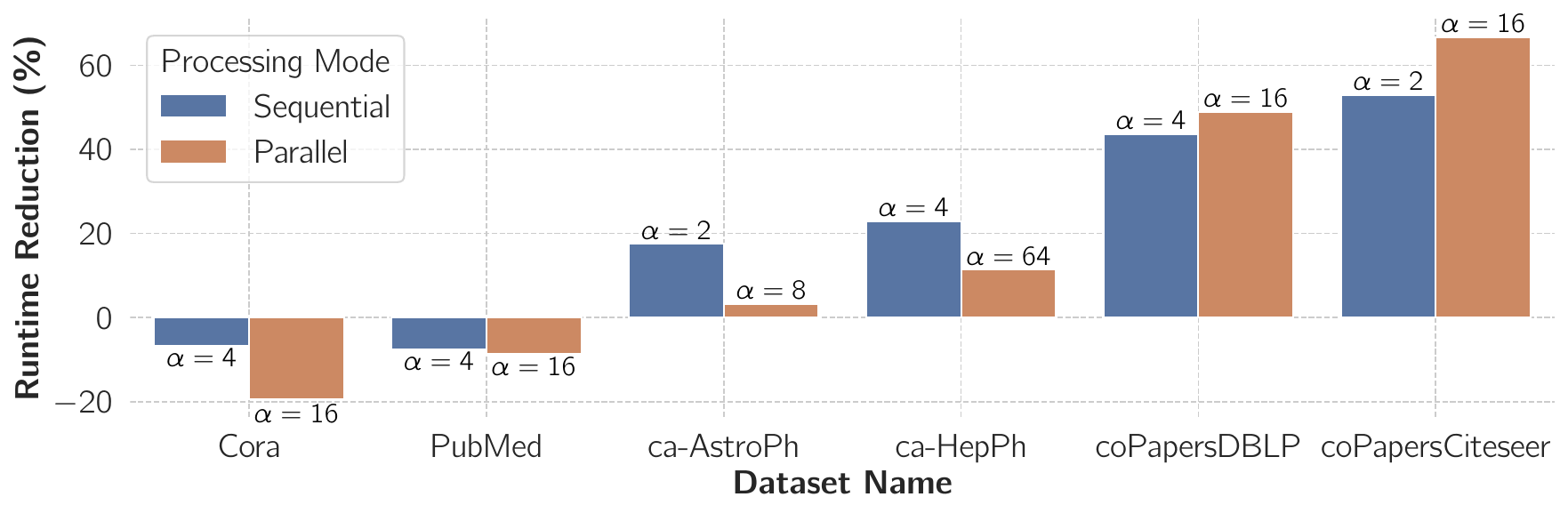}
    \caption{Performance impact of integrating the CBM format and corresponding sparse-dense matrix multiplication kernel (SpMM) in the inference stage of a 2-layer GCN across the different datasets. The y-axis shows the runtime reduction achieved in GNN inference by matrix multiplication with CBM against the matrix multiplication kernel with CSR provided by Intel MKL.
    }
    \label{fig:gcn_mkl}
\end{figure}

\subsection{Integrating CBM with GCN}\label{cbm4gcn}
To assess the impact of CBM in GNNs, we considered the runtime reduction of an inference stage of a 2-layer GCN with 500 features where the normalized adjacency matrix 
of each dataset is represent in our format. Matrix products involving the normalized adjacency matrix 
were carried out with our extended matrix multiplication kernel, as described in Section \ref{spmm_cbm}. As baseline for our experiments, we selected the same 2-layer GCN where the normalized adjacency matrix 
is represented in CSR and any products involving this matrix are carried out by the SpMM implementation found in Intel MKL.
These experiments are illustrated in Figure~\ref{fig:gcn_mkl}. To keep the discussion concise, we only considered the values of $\alpha$ that led to the best matrix multiplication performance for each dataset analyzed in Section \ref{spmm}.

Representing both Cora and PubMed datasets in CBM format increased the inference time of the corresponding GCNs compared to the baseline. This behavior is expected because the only steps of the inference that benefit from the CBM format are the products involving the normalized adjacency matrix. As previously shown, compressing these datasets using our format did not accelerate the corresponding matrix products.
Experiments with both co-author datasets demonstrated that our format can reduce the inference time for GCN models. For ca-AstroPh, our format reduced the inference time of the network by 17\% in sequential environments and 3\% in parallel environments. Similarly, for ca-HepPh, our format achieved a runtime reduction of 22\% in sequential environments and 11\% in parallel environments.
Representing both co-Paper datasets in our format resulted in the highest runtime reductions during GCN inference. For coPapersCiteseer, our format achieved an average runtime reduction of 48\% in sequential environments and 66\% in parallel environments. For coPapersDBLP, our format achieved a runtime reduction of 43\% in sequential environments and 52\% in parallel environments.

\section{Final Remarks}
In this work we proposed the Compressed Binary Matrix (CBM) format which simultaneously reduces the memory footprint of unweighted graphs and binary matrices, and enables the implementation of new matrix multiplication kernels that might be significantly faster than the current state-of-the-art.
Experimental evaluation results shown significant speedups, both in sequential and parallel environments, up to 5$\times$.
We obtained also significant performance improvements in the context of the GCN inference stage by integrating the CBM format in a deep learning framework, namely PyTorch, observing speed ups of 3$\times$.
Although we did not discuss the CBM format construction time, we observe that it can be built in a reasonable amount of time for a dataset provider. In our experiments it took us less than 16 seconds to convert the largest dataset into our format in a sequential CPU environment.
It is important to stress that the effectiveness of our format depends on the specific dataset, as discussed in the experimental evaluation. While we suspect that graphs with a high average degree and a tendency to form communities are good candidates, the best way to determine if a graph is suitable is to examine the compression ratio achieved by our format for a reasonable value of $\alpha$. Finally we highlight that our format is future proof, since future optimizations to high-performance SpMM kernels will also accelerate matrix multiplication with the CBM format.
Future work concerns integrating and evaluating the CBM format in the context of different GNNs architectures, and also targeting the training stage of this networks. Additionally, we intend to implement and evaluate our format and corresponding multiplication kernels in GPU architectures. 

\section*{Acknowledgments}
This work has been supported by the Innovation Study CBM4scale, funded by the Inno4scale project, which is funded by the European High-Performance Computing Joint Undertaking (JU) under Grant Agreement No 101118139. The JU receives support from the European Union's Horizon Europe Programme. 

\section*{Author Contributions}

\textbf{JNFA} devised the main conceptual of the Compressed Binary Matrix (CBM), including the various matrix multiplication algorithms and their optimizations. \textbf{JNFA} also implemented the CBM format and the related multiplication kernels in C++, including the integration of SpMM and AXPY (from Intel MKL) into the matrix multiplications kernels based on the CBM format, as proposed in Section~\ref{spmm_cbm}. Additionally, \textbf{JNFA} implemented an interface that enables calling the previous C++ routines from Python.
\textbf{SM} designed the Python benchmarks to compare CBM and CSR in matrix multiplication and compression quality, as described in Section~\ref{spmm}. \textbf{SM} also integrated the CBM and CSR-based matrix multiplication kernels into the Message Passing Layer (MPL) to evaluate the impact of the CBM format during the inference stage of a 2-layer GCN model implemented in PyTorch, as discussed in Section~\ref{cbm4gcn}. \textbf{SB}, \textbf{APF}, \textbf{WNG}, and \textbf{LMSR} supervised the project. \textbf{JNFA} wrote the bulk of this draft. \textbf{All authors} provided critical feedback, participated in discussions, contributed to the interpretation of the results, and approved the final manuscript.

\bibliographystyle{unsrt}  
\bibliography{references}

\end{document}